\documentclass{article}
\usepackage{amsmath,amssymb,amsfonts}
\usepackage{graphicx}
\usepackage{txfonts}

\begin{document}

\title{Number-Conserving Approaches for Atomic Bose-Einstein Condensates: An Overview}
\author{S. A. Gardiner and T. P. Billam\\
Department of Physics, Durham University, \\
Rochester Building, South Road, Durham DH1 3LE, United Kingdom}
\date{}
\maketitle

\begin{abstract}
\noindent Assuming the existence of a Bose-Einstein condensate composed of the majority of a sample of ultracold, trapped atoms, perturbative treatments to incorporate the non-condensate fraction are common.  Here we describe how this may be carried out in an explicitly number-conserving fashion, providing a common framework for the work of various authors; we also briefly consider issues of implementation, validity and application of such methods.
\end{abstract}

%\maketitle

\section{Introduction}

Since the first successful experiments in observing Bose-Einstein condensation (BEC) in cold trapped atoms \cite{anderson_ensher_95,davis_mewes_95}, there have been dramatic advances into strongly interacting (via Feshbach resonances \cite{chin_grimm_10}) and strongly correlated regimes (e.g., within optical lattices \cite{lewenstein_sanpera_07,bloch_dalibard_08}).  Nevertheless, one can still speak of a `typical' BEC experiment as consisting of a weakly interacting gas of alkali atoms, held within a confining potential formed by laser or magnetic fields, where the condensate fraction incorporates the substantial majority of constituent atoms.  In this chapter we will describe perturbative approaches, based around the existence of such a significant condensate fraction, in such a way that the many-body system is in a number eigenstate (number conserving) rather than a coherent state (symmetry breaking)
%(see also introductory discussion in Chapter \ref{Chapter_Davis_Gardiner}).  
The material presented here is intended to be a systematized amalgam of the presentations of C. W. Gardiner \cite{gardiner_97}, Castin and Dum \cite{castin_dum_98}, and of S. A. Gardiner and Morgan \cite{gardiner_morgan_07}, with some additional observations. 

We begin by contrasting so-called number-conserving approaches with the more conventional assumption of symmetry-breaking, also addressing the motivation for considering a number-conserving alternative, before describing number-conserving equivalents to the quadratic Bogolibov Hamiltonian \cite{girardeau_arnowitt_59,girardeau_98,gardiner_97,bogoliubov_47} and the linearized Bogoliubov-de Gennes equations \cite{castin_dum_98,fetter_72,degennes_book_66}.  We then cover extending this approach to a second-order minimal self-consistent treatment of dynamics, before concluding with considerations of implementation, validity and application.  Throughout, due to the preponderance of explicit time-dependences, $t$ arguments will only appear when a new quantity is introduced, as appropriate.

\section{Methodology: A Number-Conserving Perturbative Approach}

\subsection{Number-Conserving versus Symmetry-Breaking Approaches}

Our starting point is the bosonic binary interaction Hamiltonian
\begin{equation}
\hat{H} = \int d\mathbf{r} 
\hat{\Psi}^{\dagger}(\mathbf{r})
\hat{h}_{0}(\mathbf{r})
\hat{\Psi}(\mathbf{r})
+ \frac{g}{2} \int d\mathbf{r} 
\hat{\Psi}^{\dagger}(\mathbf{r}) 
\hat{\Psi}^{\dagger}(\mathbf{r})
\hat{\Psi}(\mathbf{r})
\hat{\Psi}(\mathbf{r}),
\label{gardiner_FieldHamiltonian}
\end{equation}
where the field operators obey
$[\hat{\Psi}(\mathbf{r}),
\hat{\Psi}^{\dagger}(\mathbf{r'})] = \delta(\mathbf{r}-\mathbf{r'})$,
and $\hat{h}_{0}(\mathbf{r}) = -(\hbar^2/2m) \nabla^2 + V(\mathbf{r})$;  $V(\mathbf{r})$ is an external potential, $m$ is the atomic mass, and $g = 4\pi\hbar^2a/m$, with $a$ the s-wave scattering length. The binary interaction here is characterised by a contact term $g\delta(\mathbf{r}-\mathbf{r'})$, with the conditions for which a renormalization of the consequent ultraviolet divergences is necessary described elsewhere \cite{morgan_00,morgan_04}. %(see also Chapter \ref{Chapter_Davis_Gardiner}). 
The number operator $\hat{N}=\int d\mathbf{r} \hat{\Psi}^{\dagger}(\mathbf{r})\hat{\Psi}(\mathbf{r})$ commutes with $\hat{H}$; particle number is therefore conserved, and stationary states of $\hat{H}$ must also be eigenstates of $\hat{N}$.

If (as is commonly assumed) one defines the condensate as $\phi(\mathbf{r},t) = \langle \hat{\Psi}(\mathbf{r})\rangle$ (the \textit{mean field}) \cite{pethick_smith_book_02}, the field operator can be written 
\begin{equation}
\hat{\Psi}(\mathbf{r})=\phi(\mathbf{r},t)+\hat{\psi}^{\prime}(\mathbf{r},t), 
\label{gardiner_symbreak}
\end{equation}
with fluctuations defined by $\hat{\psi}^{\prime}(\mathbf{r})= \hat{\Psi}(\mathbf{r}) - \langle \hat{\Psi}(\mathbf{r})\rangle$. A scalar function, $\phi(\mathbf{r})$ has everywhere a well-defined phase, and the  
$U(1)$ gauge symmetry of the system is \textit{broken}.  Consequently, for $\langle \hat{\Psi}(\mathbf{r})\rangle \neq 0$, the system must be in a \textit{coherent state}, i.e., a coherent superposition of $N$  \cite{fetter_walecka_book_03,castin_dum_98}.  The appropriate statistical ensemble is then grand-canonical rather than canonical, and one should work with the grand-canonical Hamiltonian $\hat{K}=\hat{H} - \mu \hat{N}$ \cite{fetter_walecka_book_03,gardiner_97}.  Assuming small fluctuations, approximate expressions may then be obtained perturbatively around purely mean-field results.  

Requiring the state of the system to be in a \textit{number\/} eigenstate implies $\langle \hat{\Psi}(\mathbf{r})\rangle = 0$.  Data is often statistically averaged from repeated experimental runs, which are unlikely to have identical shot-to-shot particle numbers --- nevertheless, if each run is for a definite (even if unknown) particle number, $ \langle \hat{\Psi}(\mathbf{r})\rangle$ is still $=0$ (one has an incoherent statistical ensemble of number eigenstates, rather than a coherent superposition).  
The alternative taken \cite{castin_dum_98,gardiner_97,girardeau_arnowitt_59,girardeau_98,morgan_00,gardiner_morgan_07} is to select a \textit{condensate mode\/} $\phi^{\mathrm{N}}(\mathbf{r},t)$:
\begin{equation}
\hat{\Psi}(\mathbf{r}) = \hat{a}_{0}(t)\phi^{\mathrm{N}}(\mathbf{r},t) + \hat{\psi}^{\mathrm{N}}(\mathbf{r},t),
\label{gardiner_orthogonal}
\end{equation}
where $\hat{\psi}^{\mathrm{N}}(\mathbf{r})$ is explicitly orthogonal to $\phi^{\mathrm{N}}(\mathbf{r})$; formally $\hat{\psi}^{\mathrm{N}}(\mathbf{r},t)=\int d\mathbf{r'}\mathcal{Q}(\mathbf{r},\mathbf{r'},t)\hat{\Psi}(\mathbf{r'})$, with the projector defined by $\mathcal{Q}(\mathbf{r},\mathbf{r'},t)=\delta(\mathbf{r}-\mathbf{r'})-\phi^{\mathrm{N}}(\mathbf{r},t)\phi^{\mathrm{N}*}(\mathbf{r'},t)$.  Following Penrose and Onsager \cite{penrose_onsager_56}, one can use the single-body density matrix $\rho(\mathbf{r},\mathbf{r'},t)=\langle \hat{\Psi}^{\dagger}(\mathbf{r'})\hat{\Psi}(\mathbf{r})\rangle$ to formally define the condensate mode as the instantaneous eigenfunction of $\rho(\mathbf{r},\mathbf{r'},t)$ with the largest eigenvalue $N_{\mathrm{c}}(t)$ (the condensate number):
\begin{equation}
\int d\mathbf{r'}\rho(\mathbf{r},\mathbf{r'},t)\phi^{\mathrm{N}}(\mathbf{r'},t) = N_{\mathrm{c}}(t)\phi^{\mathrm{N}}(\mathbf{r},t).
\label{gardiner_CondensateDefinition}
\end{equation}
It follows directly that $N_{\mathrm{c}}=\langle\hat{a}_{0}^{\dagger}\hat{a}_{0}\rangle$, and $\langle\hat{a}_{0}^{\dagger}\hat{\psi}^{\mathrm{N}}(\mathbf{r})\rangle=0$ \cite{castin_dum_98}.
Such a prescription is consistent with off-diagonal long-range order, i.e., $\rho(\mathbf{r},\mathbf{r'},t) = N_{\mathrm{c}}(t)\phi^{\mathrm{N}}(\mathbf{r},t)\phi^{\mathrm{N}*}(\mathbf{r'},t) + \langle  \hat{\psi}^{\textrm{N}\dagger}(\mathbf{r'},t)\hat{\psi}^{\mathrm{N}}(\mathbf{r},t) \rangle$ \cite{annett_book_04}.
To maintain mutual orthogonality, it is expected that nonlocal terms will arise in the dynamical equations, as indeed occurs  \cite{castin_dum_98,gardiner_morgan_07}.  The generic appearance of nonlocal terms is associated with the \textit{orthogonality\/} of condensate and non-condensate, however, not number conservation \textit{per se}.  Similarly, although the commutation relations $[\hat{\psi}^{\prime}(\mathbf{r}),\hat{\psi}^{\prime\dagger}(\mathbf{r'})]=\delta(\mathbf{r}-\mathbf{r'})$ are exactly bosonic, if the fluctuation term is defined orthogonal to the condensate mode, its commutation relations are projective: $[\hat{\psi}^{\mathrm{N}}(\mathbf{r}),\hat{\psi}^{\textrm{N}\dagger}(\mathbf{r'})]=\mathcal{Q}(\mathbf{r},\mathbf{r'})$ \cite{castin_dum_98,gardiner_97}.

\subsection{Key Concept: A Number-Conserving Fluctuation Operator}

This Chapter is concerned with describing a perturbative approach, closely analogous to treatments described in the three preceeding chapters, but leading to a theory which is guaranteed to be number-conserving by construction and does not therefore need to treat the condensate part as a mean field (see also the stochastic approaches of Part II. 2). This is achieved by introducing an appropriate operator within a number-conserving context, to analogously describe  equivalent small fluctuations to those about the mean field of a symmetry-breaking treatment. %(see Chapters \ref{Chapter_Allen_Collins}, \ref{Chapter_Walser}).

Recalling [Eq.~(\ref{gardiner_CondensateDefinition})] that $\langle\hat{a}_{0}^{\dagger}\hat{\psi}^{\mathrm{N}}(\mathbf{r})\rangle=0$, we may use $\hat{a}_{0}^{\dagger}\hat{\psi}^{\mathrm{N}}(\mathbf{r})$ as the core of an appropriate fluctuation operator; in particular, $\langle\hat{a}_{0}^{\dagger}\hat{\psi}^{\mathrm{N}}(\mathbf{r})\rangle$ is not \textit{trivially\/} $=0$, as $\hat{a}_{0}^{\dagger}\hat{\psi}^{\mathrm{N}}(\mathbf{r})$ still conserves total particle number.  Assuming an almost fully condensed system (i.e., $N_{\mathrm{c}} \approx N$), we may rescale this fluctuation term to scale approximately as $\hat{\psi}^{\mathrm{N}}(\mathbf{r})$, defining 
\begin{equation}
\hat{\Lambda}(\mathbf{r},t) = \frac{1}{\sqrt{N}}\hat{a}_{0}^{\dagger}(t)\hat{\psi}^{\mathrm{N}}(\mathbf{r},t),
\end{equation}
such that $\langle \hat{\Lambda}(\mathbf{r}) \rangle=0$ and $[\hat{\Lambda}(\mathbf{r}),\hat{\Lambda}^{\dagger}(\mathbf{r'})] \approx [\hat{\psi}^{\mathrm{N}}(\mathbf{r}),\hat{\psi}^{\textrm{N}\dagger}(\mathbf{r'})]=\mathcal{Q}(\mathbf{r},\mathbf{r'})$.
Castin and Dum define $\hat{\Lambda}(\mathbf{r},t)= \hat{N}^{-1/2}\hat{a}_{0}^{\dagger}(t)\hat{\psi}^{\mathrm{N}}(\mathbf{r},t)$; %(see also Chapter \ref{Chapter_Sinatra_Castin}); 
if the system is in a number eigenstate, and all other operators are arranged in number-conserving pairs in any approximate Hamiltonian or equation of motion, $\hat{N}$ behaves exactly as a number $N$, as made explicit  here. The  comparable $\hat{\chi}(\mathbf{r},t)$ operators of C. W. Gardiner \cite{gardiner_97} are defined through $\hat{\chi}(\mathbf{r},t) = \hat{\mathcal{N}}^{-1/2} \hat{a}_{0}^{\dagger}(t)\hat{\psi}^{\mathrm{N}}(\mathbf{r},t)$, where $\mathcal{N} = \hat{A}^{\dagger}(t)\hat{A}(t)$ and $\hat{A}(t) = \sqrt{N/N_{c}(t)}\hat{a}_{0}(t)$; hence $\langle \hat{\chi}(\mathbf{r}) \rangle$ is only approximately zero, and $[\hat{\chi}(\mathbf{r}),\hat{\chi}^{\dagger}(\mathbf{r'})] = (N_{c}/N)\mathcal{Q}(\mathbf{r},\mathbf{r'})$.  In their approximate application, to Bogoliubov order (see Section \ref{gardiner_SectionNCBogoliubov} below), $\hat{\chi}(\mathbf{r})$ and $\hat{\Lambda}(\mathbf{r})$ yield identical results.  

An equivalent operator compatible with the treatment of Girardeau and Arnowitt \cite{girardeau_arnowitt_59,girardeau_98} 
is $\hat{\Lambda}_{\mathrm{c}}(\mathbf{r},t)=\hat{N}_{\mathrm{c}}(t)^{-1/2}\hat{a}_{0}^{\dagger}(t)\hat{\psi}^{\mathrm{N}}(\mathbf{r},t)$, whereas Gardiner and Morgan use $\tilde{\Lambda}(\mathbf{r},t)=N_{\mathrm{c}}(t)^{-1/2}\hat{a}_{0}^{\dagger}(t)\hat{\psi}^{\mathrm{N}}(\mathbf{r},t)$  \cite{gardiner_morgan_07}.  The operators $\hat{\Lambda}_{\mathrm{c}}(\mathbf{r})$ and $\tilde{\Lambda}(\mathbf{r})$ are scaled by the condensate number, and hence intended to be better-tailored to describing larger non-condensate fractions. Noting the exact identities $[\hat{\Lambda}_{\mathrm{c}}(\mathbf{r}),\hat{\Lambda}_{\mathrm{c}}^{\dagger}(\mathbf{r'})]=\mathcal{Q}(\mathbf{r},\mathbf{r'})$ and $\hat{\Lambda}_{\mathrm{c}}^{\dagger}(\mathbf{r})\hat{\Lambda}_{\mathrm{c}}(\mathbf{r})=\hat{\psi}^{\textrm{N}\dagger}(\mathbf{r})\hat{\psi}^{N}(\mathbf{r})$, $\hat{\Lambda}_{\mathrm{c}}(\mathbf{r})$ may seem an attractive choice; $\langle \hat{\Lambda}_{\mathrm{c}}(\mathbf{r}) \rangle$ is not identically zero, however, whereas $\langle \tilde{\Lambda}(\mathbf{r}) \rangle =0$ is an exact identity. Hence, in the later sections of this chapter we use
$\tilde{\Lambda}(\mathbf{r})$.

\subsection{Number-Conserving Bogoliubov Treatment \label{gardiner_SectionNCBogoliubov}}

To determine an appropriate Bogoliubov Hamiltonian, while explicitly maintaining number-conservation, we first substitute Eq.\ (\ref{gardiner_orthogonal}) into Eq.\ (\ref{gardiner_FieldHamiltonian}), neglecting all terms of greater than quadratic order in $\hat{\psi}^{\mathrm{N}}(\mathbf{r})$,  $\hat{\psi}^{\mathrm{N}\dagger}(\mathbf{r})$. We set $gN=U$, and express the truncated Hamiltonian in terms of $\hat{\Lambda}(\mathbf{r})$, $\hat{\Lambda}^{\dagger}(\mathbf{r})$.  The operators $\hat{N}_{c}$, $\hat{N}_{c}(\hat{N}_{c}-1)/N$ appearing in the naively zeroth-order term in this truncated Hamiltonian must be replaced by
$N - \int d\mathbf{r'}\hat{\psi}^{\mathrm{N}\dagger}(\mathbf{r'})\hat{\psi}^{\mathrm{N}}(\mathbf{r'})\approx N - \int d\mathbf{r'}\hat{\Lambda}^{\dagger}(\mathbf{r'})\hat{\Lambda}(\mathbf{r'})$ and $N-1-2\int d\mathbf{r'}\hat{\psi}^{\mathrm{N}\dagger}(\mathbf{r'})\hat{\psi}^{\mathrm{N}}(\mathbf{r'})\approx N -1 - 2\int d\mathbf{r'}\hat{\Lambda}^{\dagger}(\mathbf{r'})\hat{\Lambda}(\mathbf{r'})$, respectively, as these include second-order terms; we also set $\hat{N}_{c}/\sqrt{N}\rightarrow \sqrt{N}$, $N/\hat{N}_{c} \rightarrow 1$, and smaller terms $\rightarrow 0$.
The consistent second-order Hamiltonian is thus \cite{gardiner_97}
\begin{equation}
\begin{split}
\hat{H}_{2} =&
H_{0}
+ \sqrt{N}
\int d\mathbf{r} 
\left\{
\phi^{\mathrm{N}*}(\mathbf{r})
\left[\hat{h}_{0}(\mathbf{r})
+U|\phi^{\mathrm{N}}(\mathbf{r})|^{2}
\right]\hat{\Lambda}(\mathbf{r})+
\mbox{H.c.}
\right\}
\\&+
\int d\mathbf{r} 
\hat{\Lambda}^{\dagger}(\mathbf{r})
\left[
\hat{h}_{0}(\mathbf{r})
+2U |\phi^{\mathrm{N}}(\mathbf{r})|^{2}
\right]
\hat{\Lambda}(\mathbf{r})
\\&
+
\frac{U}{2}\int d\mathbf{r} 
\left[
\phi^{\mathrm{N}*}(\mathbf{r})^{2}\hat{\Lambda}(\mathbf{r})^{2} + \mbox{H.c.} \right]
- \frac{U}{2}\int d\mathbf{r} |\phi^{\mathrm{N}}(\mathbf{r})|^{4}
\\& - \int d\mathbf{r'}\hat{\Lambda}^{\dagger}(\mathbf{r'})\hat{\Lambda}(\mathbf{r'})\int d\mathbf{r} \phi^{\mathrm{N}*}(\mathbf{r})\left[\hat{h}_{0}(\mathbf{r})+U|\phi^{\mathrm{N}}(\mathbf{r})|^{2}\right]\phi^{\mathrm{N}}(\mathbf{r}).
\end{split}
\label{gardiner_LambdaHamiltonianApprox}
\end{equation}
Minimizing
$H_{0}/N = 
\int d\mathbf{r} \phi^{\mathrm{N}*}(\mathbf{r})[\hat{h}_{0}(\mathbf{r})+(U/2)|\phi^{\mathrm{N}}(\mathbf{r})|^{2}]\phi^{\mathrm{N}}(\mathbf{r})$ with respect to $\phi^{\mathrm{N}}(\mathbf{r})$ (stationary, and constrained to preserve unit norm) yields the time-independent Gross-Pitaevskii equation (GPE)
\begin{equation}
\left[\hat{h}_{0}(\mathbf{r})+U|\phi^{\mathrm{N}}(\mathbf{r})|^{2}\right]\phi^{\mathrm{N}}(\mathbf{r})
= \lambda\phi^{\mathrm{N}}(\mathbf{r}),
\label{gardiner_TIGPE}
\end{equation}
where $\lambda$ arises as a Lagrange multiplier, and takes the form of a nonlinear eigenvalue \cite{gardiner_97}.  Substituting Eq.\ (\ref{gardiner_TIGPE}) back into $\hat{H}_{2}$ eliminates all terms linear in $\hat{\Lambda}(\mathbf{r})$, $\hat{\Lambda}^{\dagger}(\mathbf{r})$, leaving $\hat{H}_{2}$ in a quadratic form suitable for diagonalization by Bogoliubov transformation.

In an explicitly dynamical treatment, we use $\hat{H}_{2}$ and $[\hat{\Lambda}(\mathbf{r}),\hat{\Lambda}^{\dagger}(\mathbf{r'})] 
\approx \mathcal{Q}(\mathbf{r},\mathbf{r'})
$ 
\cite{gardiner_97,castin_dum_98,gardiner_morgan_07}
to generate the equation of motion 
$i\hbar d\hat{\Lambda}(\mathbf{r})/dt = [\hat{\Lambda}(\mathbf{r}),\hat{H}_{2}] + i\hbar \partial\hat{\Lambda}(\mathbf{r})/ \partial t$.  This yields \cite{castin_dum_98,gardiner_morgan_07}
\begin{equation}
\begin{split}
i\hbar\frac{d}{dt} \hat{\Lambda}(\mathbf{r}) =& \sqrt{N}
\int d\mathbf{r'} \mathcal{Q}(\mathbf{r},\mathbf{r'})
\left[
\hat{h}_{0}(\mathbf{r'})
+
U
 |\phi^{\mathrm{N}}(\mathbf{r'})|^{2}
 -i\hbar\frac{\partial}{\partial t}\right]
\phi^{\mathrm{N}}(\mathbf{r'})
\\
&+
\int d\mathbf{r'} 
\mathcal{Q}(\mathbf{r},\mathbf{r'})
\left[
\hat{h}_{0}(\mathbf{r'})
+
2U
 |\phi^{\mathrm{N}}(\mathbf{r'})|^{2}
 \right]
\hat{\Lambda}(\mathbf{r'})
\\&+
U\int d\mathbf{r'}
\mathcal{Q}(\mathbf{r},\mathbf{r'}) 
\hat{\Lambda}^{\dagger}(\mathbf{r'})
\phi^{\mathrm{N}}(\mathbf{r'})^{2}
-\phi(\mathbf{r})\int d\mathbf{r'}
\left[
i\hbar\frac{\partial }{\partial t}\phi^{\mathrm{N}*}(\mathbf{r'})
\right]
\hat{\Lambda}(\mathbf{r'})
\\&
-
\hat{\Lambda}(\mathbf{r})
\int d\mathbf{r'}
\phi^{\mathrm{N}*}(\mathbf{r'})
\left[
\hat{h}_{0}(\mathbf{r'})
+
U
 |\phi^{\mathrm{N}}(\mathbf{r'})|^{2}
 -i\hbar\frac{\partial}{\partial t} \right]
\phi^{\mathrm{N}}(\mathbf{r'}).
\end{split}
\label{gardiner_EOMLinearMiddle}
\end{equation}
Using $\langle \hat{\Lambda}(\mathbf{r})\rangle =0$ 
yields the time-dependent GPE 
\begin{equation}
i\hbar\frac{\partial}{\partial t}\phi^{\mathrm{N}}(\mathbf{r})=\left[\hat{h}_{0}(\mathbf{r})+U|\phi^{\mathrm{N}}(\mathbf{r})|^{2}-\lambda\right]\phi^{\mathrm{N}}(\mathbf{r}),
\end{equation}
where $\lambda=\int d\mathbf{r}\phi^{\mathrm{N}*}(\mathbf{r})\left[
\hat{h}_{0}(\mathbf{r})+U|\phi^{\mathrm{N}}(\mathbf{r})|^{2} -i\hbar\partial/\partial t
\right]\phi^{\mathrm{N}}(\mathbf{r})$ corresponds to an arbitrary global phase for $\phi^{\mathrm{N}}(\mathbf{r})$ \cite{castin_dum_98}.\footnote{Note that we regain the time-independent GPE for $i\hbar\partial\phi^{\mathrm{N}}(\mathbf{r})/\partial t = 0$.} 
Substituting this result back into Eq.\ (\ref{gardiner_EOMLinearMiddle}) simplifies the expression, which can be combined with its Hermitian conjugate to form \cite{castin_dum_98,gardiner_morgan_07}
\begin{equation}
\begin{split}
i\hbar\frac{d}{dt}
\left(
\begin{array}{c}
\hat{\Lambda}(\mathbf{r}) \\
\hat{\Lambda}^{\dagger}(\mathbf{r})
\end{array}
\right) =& \int d\mathbf{r'} 
\mathcal{L}(\mathbf{r},\mathbf{r'})
\left(
\begin{array}{c}
\hat{\Lambda}(\mathbf{r'}) \\
\hat{\Lambda}^{\dagger}(\mathbf{r'})
\end{array}
\right)
\\
\mathcal{L}(\mathbf{r},\mathbf{r'})=&
\left(
\begin{array}{cc}
L(\mathbf{r},\mathbf{r'}) & M(\mathbf{r},\mathbf{r'}) \\
-M^{*}(\mathbf{r},\mathbf{r'}) & -L^{*}(\mathbf{r},\mathbf{r'})
\end{array}
\right),
\\
L(\mathbf{r},\mathbf{r'})= &
\delta(\mathbf{r}-\mathbf{r'})[\hat{h}_{0}(\mathbf{r'}) + U
|\phi^{\mathrm{N}}(\mathbf{r'})|^2 -\lambda] 
\\&
+U
\int d\mathbf{r'} \mathcal{Q} (\mathbf{r},\mathbf{r'}) 
|\phi(\mathbf{r'})^{\mathrm{N}}|^2  \mathcal{Q} (\mathbf{r'},\mathbf{r'}),
\\
M(\mathbf{r},\mathbf{r'})= &
U
\int d\mathbf{r'} \mathcal{Q} (\mathbf{r},\mathbf{r'}) 
\phi^{\mathrm{N}}(\mathbf{r'})^2  \mathcal{Q}^{*} (\mathbf{r'},\mathbf{r'}),
\end{split}
\label{gardiner_MBDGE}
\end{equation}
i.e., the standard Bogoliubov-de Gennes equations (BDGE) modified only by the appearance of nonlocal projector terms.  Note that the treatment of C. W. Gardiner \cite{gardiner_97} avoids employing $\langle \hat{\Lambda}(\mathbf{r}) \rangle =0$, instead transforming to a `condensate picture.'  Use of this identity is well-motivated, however, and yields equivalent results in a simpler fashion.

The $\mathcal{L}(\mathbf{r},\mathbf{r'})$ can now be diagonalized \cite{castin_dum_98,gardiner_morgan_07}:
\begin{equation}
\mathcal{L}(\mathbf{r},\mathbf{r'}) =
\sum_{k=1}^{\infty}
\epsilon_{k} 
\left[
\left(
\begin{array}{c}
u_{k}(\mathbf{r})
\\
v_{k}(\mathbf{r})
\end{array}
\right)
(
u_{k}^{*}(\mathbf{r'}), -v_{k}^{*}(\mathbf{r'})
)
-
\left(
\begin{array}{c}
v_{k}^{*}(\mathbf{r})
\\
u_{k}^{*}(\mathbf{r})
\end{array}
\right)
(
-v_{k}(\mathbf{r'}), u_{k}(\mathbf{r'})
)
\right],
\label{gardiner_SpectralDecomp}
\end{equation}
($k$ is an arbitrary index) where $(\phi^{\mathrm{N}}(\mathbf{r}),0)$ or $(0,\phi^{\mathrm{N}*}(\mathbf{r}))$ are also eigenstates, with eigenvalue $=0$. The spectrum $\epsilon_{k}$ is identical to that associated with the BDGE, and the eigenstates differ only in that they are explicitly orthogonal to the condensate mode \cite{castin_dum_98}.
We decompose $(\hat{\Lambda}(\mathbf{r}),\hat{\Lambda}^{\dagger}(\mathbf{r}))$ as
\begin{equation}
\left(
\begin{array}{c}
\hat{\Lambda}(\mathbf{r})
\\
\hat{\Lambda}^{\dagger}(\mathbf{r})
\end{array}
\right) 
=
\sum_{k=1}^{\infty}
\left[
\hat{b}_{k}
\left(
\begin{array}{c}
u_{k}(\mathbf{r})
\\
v_{k}(\mathbf{r})
\end{array}
\right)
+
\hat{b}_{k}^{\dagger}
\left(
\begin{array}{c}
v_{k}^{*}(\mathbf{r})
\\
u_{k}^{*}(\mathbf{r})
\end{array}
\right)\right];
\label{gardiner_LDecompMain}
\end{equation}
the orthonormality relations 
$\int d\mathbf{r}
[u_{k'}^{*}(\mathbf{r})u_{k}(\mathbf{r})
-
v_{k'}^{*}(\mathbf{r})v_{k}(\mathbf{r})]
=  \delta_{kk'}$, 
$\int d\mathbf{r}
[u_{k'}(\mathbf{r})v_{k}(\mathbf{r})
-
v_{k'}(\mathbf{r})u_{k}(\mathbf{r})]
= 0$, apply, meaning that the quasiparticle annihilation operators
$
\hat{b}_{k} = \int d\mathbf{r} 
[ u_{k}^{*}(\mathbf{r})\hat{\Lambda}(\mathbf{r})
-
v_{k}^{*}(\mathbf{r})\hat{\Lambda}^{\dagger}(\mathbf{r}) ]
$ with their Hermitian conjugates form a bosonic algebra, i.e., $[\hat{b}_{k},\hat{b}_{k'}^{\dagger}] = \delta_{kk'}$ (as $[\hat{\Lambda}(\mathbf{r}),
\hat{\Lambda}^{\dagger}(\mathbf{r'})]\approx \mathcal{Q}(\mathbf{r},\mathbf{r'})$ to current order) \cite{castin_dum_98,gardiner_morgan_07}. The system dynamics are to this order described by the modified BDGE coupled to the time-dependent GPE. This can lead to a situation of rapid non-condensate growth, but condensate depletion is not accounted for \cite{gardiner_jaksch_00,reslen_creffield_08}, which effectively describes a zero-temperature, infinite particle limit \cite{gardiner_morgan_07}; a treatment to the same order using $N_{\mathrm{c}}^{-1/2}$
 (rather than $N^{-1/2}$) as an asymptotic expansion parameter \cite{morgan_00,gardiner_morgan_07} is therefore functionally equivalent, as $N_{\mathrm{c}}$ must $\approx N$.
Assuming a stationary configuration (such that first-order terms are eliminated), $\hat{H}_{2}$ can now be written in diagonal form:
\begin{equation}
\hat{H}_{2} = H_{0} -
\frac{U}{2}\int d\mathbf{r} |\phi^{\mathrm{N}}(\mathbf{r})|^{4}
+\sum_{k=1}^{\infty}
\epsilon_{k}\left[
\hat{b}_{k}^{\dagger}\hat{b}_{k}
-\int d\mathbf{r} |v_{k}(\mathbf{r})|^{2}\right],
\end{equation}
i.e., we have carried out an equivalent, number-conserving Bogoliubov transformation.

\subsection{Second-Order Self-Consistent Treatment}
To account for significant thermal or dynamical depletion, we must go to higher order. Using second-order perturbation theory,%(yielding a number-conserving parallel to the approaches of Chapters \ref{Chapter_Allen_Collins}, \ref{Chapter_Walser}, \ref{Chapter_Hanna_MurPetit}),
\footnote{Note, however, Morgan's remark that Hartree-Fock-Bogoliubov factorizations of third- and fourth-order terms neglect third-order corrections as large as fourth-order terms that are retained \cite{morgan_00}.} Morgan determined a fourth-order approximate Hamiltonian,  for a system at thermal equilibrium \cite{morgan_00}, which had a gapless excitation spectrum, in accordance with the Hugenholtz-Pines theorem \cite{hugenholtz_pines_59}.  Here we briefly describe the dynamical second-order treatment of S. A. Gardiner and Morgan  \cite{gardiner_morgan_07}, in terms of $\tilde{\Lambda}(\mathbf{r})$; this is the minimal order necessary for the consistent treatment of particle transfer between condensate and non-condensate. 

The first term in 
$
i\hbar d\tilde{\Lambda}(\mathbf{r})/dt =
\left(i\hbar d N_{\mathrm{c}}/dt\right)\tilde{\Lambda}(\mathbf{r})/2N_{\mathrm{c}}  + 
N_{\mathrm{c}}^{-1/2}
(i\hbar d
[
\hat{a}_{0}^{\dagger}\hat{\psi}^{\mathrm{N}}(\mathbf{r})
]/dt)
$
is of cubic order, and is dropped \cite{gardiner_morgan_07}. One can substitute in the first 3 terms of Eq.\ (A4) in \cite{castin_dum_98}, and express the result in terms of $\tilde{\Lambda}(\mathbf{r},t)$, $\tilde{\Lambda}^{\dagger}(\mathbf{r},t)$, $\tilde{U}=gN_{\mathrm{c}}$, setting $\hat{N}_{c}/N_{\mathrm{c}}\rightarrow 1$, $\sqrt{N_{\mathrm{c}}}/\hat{N}_{c}\rightarrow 1/\sqrt{N_{\mathrm{c}}}$ and smaller terms to zero.  Taking the expectation value then yields the generalized GPE (GGPE) 
\begin{equation}
\begin{split}
i\hbar\frac{\partial}{\partial t} \phi^{\mathrm{N}}(\mathbf{r})=&\left[
H_{\mathrm{g}}(\mathbf{r})
-\lambda_{2} 
\right]
\phi(\mathbf{r})
+
\tilde{U}\phi^{\mathrm{N}*}(\mathbf{r})
\frac{\langle
\tilde{\Lambda}(\mathbf{r})^{2}
\rangle}{N_{\mathrm{c}}}
-
\frac{\tilde{U}}{N_{\mathrm{c}}}
\int d\mathbf{r'}
 |\phi^{\mathrm{N}}(\mathbf{r'})|^{2}
\\& \times
\left[
\langle
\tilde{\Lambda}^{\dagger}(\mathbf{r'})\tilde{\Lambda}(\mathbf{r})
\rangle
\phi^{\mathrm{N}}(\mathbf{r'})
+
\phi^{\mathrm{N}*}(\mathbf{r'})
\langle
\tilde{\Lambda}(\mathbf{r'})\tilde{\Lambda}(\mathbf{r})
\rangle
\right],
\end{split}
\label{gardiner_GeneralizedGrossPitaevskiiEnd}
\end{equation}
where
\begin{equation}
\begin{split}
H_{\mathrm{g}}(\mathbf{r}) =&
\hat{h}_{0}(\mathbf{r})
+
\tilde{U}
\left[
\left(
1-\frac{1}{N_{\mathrm{c}}}
\right)
 |\phi^{\mathrm{N}}(\mathbf{r})|^{2}
+
2
\frac{\langle
\tilde{\Lambda}^{\dagger}(\mathbf{r})\tilde{\Lambda}(\mathbf{r})
\rangle}{N_{\mathrm{c}}} \right],
\\
\lambda_{2} = &\int d\mathbf{r}
\left\{
\phi^{\mathrm{N}*}(\mathbf{r})
\left[
H_{\mathrm{g}}(\mathbf{r})
 -i\hbar\frac{\partial}{\partial t}
 \right]
\phi^{\mathrm{N}}(\mathbf{r})
+
\tilde{U}
\phi^{\mathrm{N}*}(\mathbf{r})^{2}
\frac{\langle
\tilde{\Lambda}(\mathbf{r})^{2}
\rangle}{N_{\mathrm{c}}}
 \right\} \;.
\end{split}
\end{equation}
This is effectively equivalent to the GPE in combination with a separate second-order correction \cite{castin_dum_98}; %(see also Chapter \ref{Sinatra_Castin}); 
the nonlinearity of the equations of motion prevents the expression of such a correction in closed form, however. Note the nonlocal terms (off-diagonal forms of the normal and anomolous average), as well as the ultraviolet-divergent diagonal anomalous average, which must be renormalized \cite{morgan_00,morgan_04}.  The GGPE can also be generated from the approximate cubic Hamiltonian
\begin{equation}
\begin{split}
\hat{H}_{3} =& \hat{H}_{2}^{N_{\mathrm{c}}} 
+
\lambda_{0}
\int d\mathbf{r'}
\langle \tilde{\Lambda}^{\dagger}(\mathbf{r'})
\tilde{\Lambda}(\mathbf{r'})\rangle 
\\&+
\frac{\tilde{U}}{\sqrt{N_{\mathrm{c}}}}
\int d\mathbf{r} 
\Bigl\{
\phi^{\mathrm{N}*}(\mathbf{r})
\left[
2\langle\tilde{\Lambda}^{\dagger}(\mathbf{r})\tilde{\Lambda}(\mathbf{r})\rangle\tilde{\Lambda}(\mathbf{r})
+
\tilde{\Lambda}^{\dagger}(\mathbf{r})\langle\tilde{\Lambda}(\mathbf{r})^{2}\rangle
\right]
+\mbox{H.c.}
\Bigr\}
\\&
-\frac{\tilde{U}}{\sqrt{N_{\mathrm{c}}}}
\int d\mathbf{r}
\left[
\phi^{\mathrm{N}*}(\mathbf{r})|\phi^{\mathrm{N}}(\mathbf{r})|^{2}\tilde{\Lambda}(\mathbf{r}) 
+
\mbox{H.c.}
\right]
\\&
+
\frac{\tilde{U}}{\sqrt{N_{\mathrm{c}}}}
\iint d\mathbf{r}d\mathbf{r'} 
\Bigl\{
\phi^{\mathrm{N}*}(\mathbf{r})|\phi^{\mathrm{N}}(\mathbf{r})|^{2} 
\\&\times
\left[
\langle
\tilde{\Lambda}^{\dagger}(\mathbf{r'}) 
\tilde{\Lambda}(\mathbf{r})
\rangle
\tilde{\Lambda}(\mathbf{r'}) 
+
\tilde{\Lambda}^{\dagger}(\mathbf{r'}) 
\langle
\tilde{\Lambda}(\mathbf{r'})
\tilde{\Lambda}(\mathbf{r}) 
\rangle
\right]
+\mbox{H.c.}
\Bigr\},
\end{split}
\label{gardiner_LambdaTildeHamiltonianThirdOrderGaussian}
\end{equation}
in conjunction with the more complete form of the commutator
$[\tilde{\Lambda}(\mathbf{r}),\tilde{\Lambda}^{\dagger}(\mathbf{r'})] \approx
\mathcal{Q}(\mathbf{r},\mathbf{r'})  -
\langle \tilde{\Lambda}^{\dagger}(\mathbf{r'})\tilde{\Lambda}(\mathbf{r}) \rangle/N_{\mathrm{c}}
$. Both assume a Gaussian approximation  \cite{gardiner_morgan_07}, and $\hat{H}_{2}^{N_{\mathrm{c}}}$, $\lambda_{0}$ are equivalent to $\hat{H}_{2}$ of Eq.\ (\ref{gardiner_LambdaHamiltonianApprox}), $\lambda$, with $N$, $U$ replaced by $N_{\mathrm{c}}$, $\tilde{U}$. 

Substituting the GGPE back into the second-order equations of motion for $\tilde{\Lambda}(\mathbf{r})$, $\tilde{\Lambda}^{\dagger}(\mathbf{r})$ generated by $\hat{H}_{3}$ eliminates all zeroth- and second-order terms: what remain (discarding higher-order terms) are the modified BDGE [Eq. (\ref{gardiner_MBDGE})], with $U$, $N$, $\hat{\Lambda}(\mathbf{r})$ replaced by $\tilde{U}$, $N_{\mathrm{c}}$, $\tilde{\Lambda}(\mathbf{r})$.  Similarly,  substituting the time-independent GGPE back into $\hat{H}_{3}$ yields
\begin{equation}
\begin{split}
\hat{H}_{3} =& H_{0}^{N_{\mathrm{c}}}
+
\lambda_{0}\int d\mathbf{r}
\langle \tilde{\Lambda}^{\dagger}(\mathbf{r})
\tilde{\Lambda}(\mathbf{r})\rangle
-
\frac{\tilde{U}}{2}
\int d\mathbf{r}
 |\phi^{\mathrm{N}}(\mathbf{r})|^{4}
\\
&+
\int d\mathbf{r} 
\tilde{\Lambda}^{\dagger}(\mathbf{r})
\left[
\hat{h}_{0}(\mathbf{r})
+
2\tilde{U}
 |\phi^{\mathrm{N}}(\mathbf{r})|^{2}
-\lambda_{0} \right]
\tilde{\Lambda}(\mathbf{r})
\\&
+
\frac{\tilde{U}}{2}\int d\mathbf{r} 
\left[
\phi^{\mathrm{N}*}(\mathbf{r})^{2}\tilde{\Lambda}(\mathbf{r})^{2} +
\mbox{H.c.}
\right],
\end{split}
\label{gardiner_LambdaTildeHamiltonianSecondOrderGaussianTimeIndependent}
\end{equation}
where $H_{0}^{N_{\mathrm{c}}}$, is equivalent to $H_{0}$, with $N$, $U$ replaced by $N_{\mathrm{c}}$, $\tilde{U}$, and we note that the first corrections to $H_{0}=H_{0}^{N_{\mathrm{c}}}+\lambda_{0}\int d\mathbf{r}
\langle \tilde{\Lambda}^{\dagger}(\mathbf{r})
\tilde{\Lambda}(\mathbf{r})\rangle$ appear at quartic order; other corrections to $\tilde{U}=U$ are smaller than need be accounted for. $\hat{H}_{3}$ and $\hat{H}_{2}$ are therefore effectively identical for a stationary state, and $\hat{H}_{3}$ is literally identical to the stationary form of the second-order approximate Hamiltonian determined from an asymptotic expansion in powers of $N_{\mathrm{c}}^{-1/2}$ \cite{gardiner_morgan_07} (rather than $N^{-1/2}$ \cite{gardiner_97,castin_dum_98}).  In terms of quasiparticles $\tilde{b}_{k}$, $\tilde{b}_{k}^{\dagger}$ [defined with respect to $\tilde{\Lambda}(\mathbf{r})$, $\tilde{\Lambda}^{\dagger}(\mathbf{r})$ as $\hat{b}_{k}$, $\hat{b}_{k}^{\dagger}$ are defined with respect to $\hat{\Lambda}(\mathbf{r})$, $\hat{\Lambda}^{\dagger}(\mathbf{r})$] this becomes (assuming a thermal equilibrium state and bosonic quasiparticles):
\begin{equation}
\begin{split}
\hat{H}_{3}
= &
H_{0}^{N_{\mathrm{c}}}
-
\frac{\tilde{U}}{2}\int d\mathbf{r} |\phi^{\mathrm{N}}(\mathbf{r})|^{4}
+\sum_{k=1}^{\infty} \epsilon_{k} 
\left[
\tilde{b}_{k}^{\dagger}\tilde{b}_{k}
-
\int d\mathbf{r} |v_{k}(\mathbf{r})|^{2}
\right]
\\&
+
\lambda_{0}
\sum_{k=1}^{\infty}\int d\mathbf{r}
\left\{
(\langle
\tilde{b}_{k}^{\dagger}\tilde{b}_{k}
\rangle + 1)
|v_{k}(\mathbf{r})|^{2}
+
\langle
\tilde{b}_{k}^{\dagger}\tilde{b}_{k}
\rangle
[
|v_{k}(\mathbf{r})|^{2}+1
]\right\}.
\end{split}
\label{Eq:ExtendedEnergyFunctionalEquilibrium}
\end{equation}
The significance of deviations from exactly bosonic quasiparticle commutation relations with increasing depletion is difficult to quantify in the abstract \cite{gardiner_morgan_07}; we note that this does not seem to have been an issue in the finite-temperature calculations of Morgan, however  \cite{morgan_rusch_03,morgan_04,morgan_05}.

The $\lambda_{2}$ is in general complex; this constrains $\phi^{\mathrm{N}}(\mathbf{r})$ to unit norm 
even though the GGPE includes terms describing particle transfer between condensate and non-condensate. It may be advantageous to consider a condensate wavefunction with varying norm: $\Phi(\mathbf{r}) = \sqrt{N_{\mathrm{c}}/N}\phi^{\mathrm{N}}(\mathbf{r})$.  Noting that $N_{\mathrm{c}}=N-\int d\mathbf{r}\langle \tilde{\Lambda}^{\dagger}(\mathbf{r})\tilde{\Lambda}(\mathbf{r}) \rangle$, condensate number dynamics can be written (to quadratic order) as
$
i\hbar dN_{c}/dt 
= \tilde{U}\int d\mathbf{r} 
[
\phi^{\mathrm{N}*}(\mathbf{r})^{2}
\langle \tilde{\Lambda}(\mathbf{r})^{2}\rangle 
-\langle\tilde{\Lambda}^{\dagger}(\mathbf{r})^{2}\rangle
\phi^{\mathrm{N}}(\mathbf{r})^{2}
] = (\lambda_{2}-\lambda_{2}^{*})N_{\mathrm{c}}
$ \cite{gardiner_morgan_07}.
Hence, from $i\hbar \partial \Phi(\mathbf{r})/\partial t = \sqrt{N_{\mathrm{c}}/N}[ i\hbar \partial \phi^{\mathrm{N}}(\mathbf{r})/\partial t] +(i\hbar dN_{c}/dt)\Phi(\mathbf{r})/2N_{\mathrm{c}}$,
\begin{equation}
\begin{split}
i\hbar\frac{\partial}{\partial t} \Phi(\mathbf{r})=&\left[
H_{\mathrm{g}}(\mathbf{r})
-\lambda_{\mathrm{R}} 
\right]
\Phi(\mathbf{r})
+
U\Phi^{*}(\mathbf{r})
\frac{\langle
\tilde{\Lambda}(\mathbf{r})^{2}
\rangle}{N}
-
\frac{U}{N_{\mathrm{c}}}
\int d\mathbf{r'}
 |\Phi(\mathbf{r'})|^{2}
\\& \times
\left[
\langle
\tilde{\Lambda}^{\dagger}(\mathbf{r'})\tilde{\Lambda}(\mathbf{r})
\rangle
\Phi(\mathbf{r'})
+
\Phi^{*}(\mathbf{r'})
\langle
\tilde{\Lambda}(\mathbf{r'})\tilde{\Lambda}(\mathbf{r})
\rangle
\right],
\end{split}
\label{gardiner_xiGeneralizedGrossPitaevskiiEnd}
\end{equation}
where
\begin{equation}
\begin{split}
H_{\mathrm{g}}(\mathbf{r}) =&
\hat{h}_{0}(\mathbf{r})
+
U
\left[
\left(
1-\frac{1}{N_{\mathrm{c}}}
\right)
 |\Phi(\mathbf{r})|^{2}
+
2
\frac{\langle
\tilde{\Lambda}^{\dagger}(\mathbf{r})\tilde{\Lambda}(\mathbf{r})
\rangle}{N} \right],
\\
\lambda_{\mathrm{R}} = &
\frac{N}{N_{\mathrm{c}}}
\int d\mathbf{r}
\Phi^{*}(\mathbf{r})
\left[
H_{\mathrm{g}}(\mathbf{r})
 -i\hbar\frac{\partial}{\partial t}
 \right]
\Phi(\mathbf{r})
+
\frac{U}{2N_{\mathrm{c}}}
\int d\mathbf{r}
\left[
\Phi^{*}(\mathbf{r})^{2}
\langle
\tilde{\Lambda}(\mathbf{r})^{2}
\rangle
+ \mbox{H.c.}
\right].
\end{split}
\end{equation}
Here $\lambda_{\mathrm{R}}$ is real, and so, like $\lambda$, simply describes an arbitrary phase. Carrying this formulation over to the modified BDGE, we note that, e.g., $\tilde{U}|\phi^{\mathrm{N}}(\mathbf{r})|^{2} = U|\Phi(\mathbf{r})|^{2}$. Hence, the correct equations are obtained by setting $\phi^{\mathrm{N}}(\mathbf{r}) \rightarrow \Phi(\mathbf{r})$ in Eq.\ (\ref{gardiner_MBDGE}).  Another advantage is that the condensate number can be tracked by  $N_{\mathrm{c}}=\int d\mathbf{r} |\Phi(\mathbf{r})|^{2}$, rather than the more involved procedure of continuously determining $N_{\mathrm{c}}=N- \int d\mathbf{r}\langle \tilde{\Lambda}^{\dagger}(\mathbf{r})\tilde{\Lambda}(\mathbf{r}) \rangle$.

\subsection{Numerical Implementation}
Within the number conserving Bogoliubov (i.e., dynamically first-order) treatment, first determine an appropriate stationary solution $\phi^{\mathrm{N}}(\mathbf{r})$ of the time-independent GPE, with its corresponding $\lambda$.  With these, construct and diagonalize $\mathcal{L}(\mathbf{r},\mathbf{r'})$ to determine its eigenmodes. Choosing the quasiparticle operators to be time-independent \cite{castin_dum_98}, the dynamics of $\hat{\Lambda}(\mathbf{r})$, $\hat{\Lambda}^{\dagger}(\mathbf{r})$ are entirely determined by
\begin{equation}
i\hbar\frac{\partial}{\partial t}
\left(
\begin{array}{c}
u_{k}(\mathbf{r}) \\
v_{k}(\mathbf{r})
\end{array}
\right) = \int d\mathbf{r'} 
\mathcal{L}(\mathbf{r},\mathbf{r'})
\left(
\begin{array}{c}
u_{k}(\mathbf{r'}) \\
v_{k}(\mathbf{r'})
\end{array}
\right),
\end{equation}
propagating as many initial eigenmodes as deemed necessary.  During the calculation, one can discard the projectors $\mathcal{Q}(\mathbf{r},\mathbf{r'})$, $\mathcal{Q}^{*}(\mathbf{r},\mathbf{r'})$, regaining the standard BDGE.  Applying 
\begin{equation}
\left(
\begin{array}{cc}
\mathcal{Q}(\mathbf{r},\mathbf{r'}) & 0 \\
0 & \mathcal{Q}^{*}(\mathbf{r},\mathbf{r'})
\end{array}
\right)
\end{equation}
to the final propagated modes then gives the correct result \cite{castin_dum_98}.  One must propagate the time-dependent GPE (with $\lambda$ set $=0$) in parallel, feeding the solution into the BDGE (and, ultimately, the projectors). Nonlocality of the modified BDGE is in this case not a practical issue.  

Determining a stationary GPE solution is also the starting point when applying the second-order treatment. To study dynamics starting from an ultracold sample with initially negligible depletion, the initial setup should be equivalent, except that one must determine the normal and anomalous average terms to feed into the GGPE (note that the anomalous terms in particular are generally slow to converge, although semiclassical approximations can significantly help \cite{morgan_04,morgan_05}).  If the dynamics induce substantial depletion, the GGPE evolution will soon differ from that produced by the GPE.  In the modified BDGE to which the GGPE is coupled, the projectors do not now separate out; hence, nonlocality due to the projectors and the off-diagonal normal and anomalous terms is a significant numerical issue.  A finite temperature initial condition must be produced from an initial stationary GPE solution in an iterative, self-consistent manner \cite{morgan_00,morgan_thesis_99}.

\section{Validity Issues}

The treatments presented here are effectively perturbative expansions of increasing orders of fluctuation terms about a classical field (the condensate). As such, their validity relies upon  $(Na^{3})^{1/2}\ll 1$ if $T=0$, and $(k_{\mathrm{B}}T/gN_{\mathrm{c}})(N_{\mathrm{c}}a^{3})^{1/2} \ll 1$ if $(k_{\mathrm{B}}T/gN_{\mathrm{c}})\gg 1$, where $T$ is the temperature and $k_{\mathrm{B}}$ is Boltzmann's constant \cite{morgan_00}.  More stringently, dynamics propagated by the GPE coupled to the modified BDGE are only valid so long as the the non-condensate fraction [as determined by $\int d\mathbf{r} \langle \hat{\Lambda}^{\dagger}(\mathbf{r})\hat{\Lambda}(\mathbf{r})\rangle$] remains insignificant, i.e., $1-N_{\mathrm{c}}/N \ll 1$.  

The second-order Gardiner--Morgan treatment \cite{gardiner_morgan_07} is the basis of Morgan's analysis of excitations to finite-temperature BEC \cite{morgan_rusch_03,morgan_04,morgan_05}, to good agreement with experiment \cite{jin_ensher_96}.  Good agreement was also achieved also achieved by the ZNG treatment \cite{zaremba_nikuni_99} %(described in Chapter \ref{Chapter_Allen_Collins}) 
of Jackson and Zaremba \cite{jackson_zaremba_02}.  We note this treatment (unlike Gardiner--Morgan \cite{gardiner_morgan_07}) does not appear to explicitly account for the phonon character of low-energy states or the anomalous average and Beliaev processes \cite{morgan_rusch_03}, which can be significant \cite{morgan_00,hutchinson_dodd_98,bretin_rosenbusch_03,katz_steinhauer_02,mizushima_ichioka_03}.  In summary, ZNG seems more consciously oriented towards regimes of thermal equilibrium (and as such is numerically more tractable), whereas the approach presented here includes effects that are likely to be more significant at low initial temperatures or situations far from equilibrium.  

\begin{figure}[t]
\centerline{\includegraphics[width=11cm]{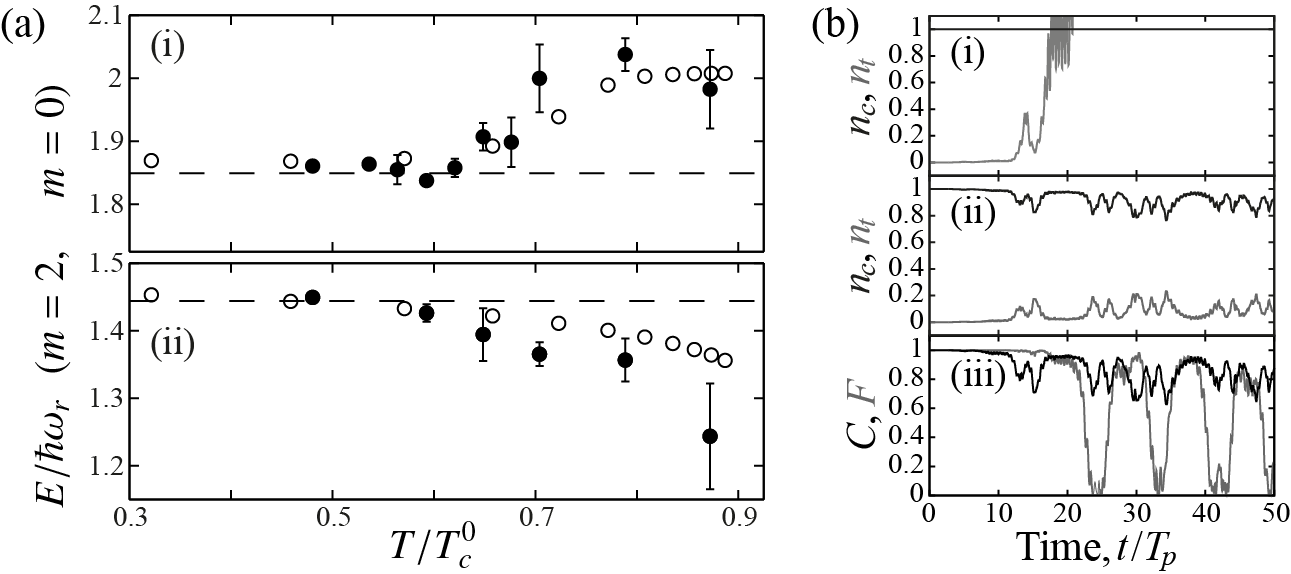}} 
\caption{Implementations of the second-order self-consistent number-conserving approach to Bose--Einstein condensate and non-condensate dynamics developed by Gardiner and Morgan \cite{gardiner_morgan_07}.
(a) Figure adapted from \cite{morgan_rusch_03} comparing experimental \cite{jin_ensher_96} (solid circles) and theoretical (open circles) excitation energies for excitation modes with axial quantum numbers (i) $m=0$ and (ii) $m=2$ as a function of reduced temperature $T/T_{c}^{0}$.  Here $T_{c}^{0}$ is the BEC critical temperature for an ideal gas, and $\omega_{r}$ is the radial trapping frequency.
(b) Figure adapted from \cite{billam_gardiner_11}, of sample dynamics of a $\delta$-kicked-rotor BEC, a simplified spatially-periodic quasi-1d system with time-periodic driving potential $V(\theta,t)=\kappa\cos(\theta)\sum_{j=0}^{\infty}\delta(t-jT_{p})$, showing condensate $n_{c}$ and non-condensate $n_{t}$ fractions evolved by (i) first-order and (ii) second-order number-conserving descriptions. (b)(iii) plots of a coherence measure $C=\iint d\theta d\theta' g_{1}(\theta,\theta')g_{1}(\theta',\theta)$, where $g_{1}(\theta,\theta') = \langle \hat{\Psi}^{\dagger}(\theta')\hat{\Psi}(\theta)\rangle/N$, and the fidelity of the condensate wave function to the GPE evolution $F= |\int d\theta \phi_{\textrm{GPE}}^{*}\phi(\theta)|^{2}$ for dynamics corresponding to (b)(ii).
} 
\label{gardiner_fig1} 
\end{figure}

Another contrasting approach able to account for a significant non-condensate fraction is that based on a cumulant expansion developed by K\"{o}hler \cite{kohler_burnett_02} and coworkers. %(see Chapter \ref{Hanna_Mur-Petit}).  
This has been very successful in describing the formation of Feshbach molecules, and also accounts for the dynamical loss of (non-molecule) atoms from the initial condensate fraction \cite{kohler_gasenzer_03}.  An essential difference in this approach is the inclusion of a more physical scattering potential capable of supporting bound states (molecules), consequently also avoiding renormalization issues.  As such, significant condensate depletion can occur due essentially to pure two-body, relatively high-energy scattering processes, without necessarily addressing loss due to the low-energy, many-body processes recently studied by Billam and Gardiner \cite{billam_gardiner_11}, and others \cite{gardiner_jaksch_00,castin_dum_97,zhang_liu_04,liu_zhang_06,reslen_creffield_08,monteiro_rancon_09}.

\section{Applications}

Morgan's analysis of finite-temperature BEC excitations \cite{morgan_rusch_03,morgan_04,morgan_05} [see Fig.~\ref{gardiner_fig1}(a)]  involved applying a linear response treatment to the second-order equations, justified by the excitations being due to small perturbations; the equations take a rather involved appearance as a result, but are numerically more tractable than a full dynamical calculation.  A fully dynamical treatment, however, is be necessary to study non-perturbative dynamics \cite{gardiner_jaksch_00, reslen_creffield_08,monteiro_rancon_09} where significant depletion from an initial very low temperature Bose-Einstein condensate is expected, self consistently.  This has recently been carried out by Billam and Gardiner within a simple quasi-1d $\delta$-kicked-rotor-BEC configuration [see Fig.~\ref{gardiner_fig1}(b)].  The kicked rotor is well known system in the context of chaotic and quantum chaotic dynamics, and has had numerous atom-optical realizations \cite{moore_robinson_95,oberthaler_godun_99,duffy_mellish_04,ryu_anderson_06}, as well as being an ideal dynamical test-system.  An important conclusion is that a BEC appears to be impressively robust, in contrast to what might be expected from first-order treatments \cite{gardiner_jaksch_00,zhang_liu_04,liu_zhang_06,reslen_creffield_08}.

\section*{Acknowledgments}

SAG would like to thank P. M. Sutcliffe for useful discussions; we are also grateful for the support of the UK EPSRC (Grant No.\ EP/G056781/1), and TPB for that of Durham University.

%\newpage

%\bibliographystyle{ws-rv-van}
%%\bibliography{Gardiner.bib}
%\bibliography{FINESS_Book_New.bib}

\end{document}